\documentclass[aps,preprint,showpacs,preprintnumbers,amsmath,amssymb]{revtex4}
\usepackage[dvips]{graphicx}
\usepackage{float}
\begin{document}

\title{Perturbative approach to the problem of particle production in electric field on de Sitter universe}
\author{Cosmin Crucean \thanks{E-mail:~~crucean@physics.uvt.ro}
and Mihaela-Andreea B\u aloi \thanks{E-mail:~~mihaela.baloi88@e-uvt.ro}\\
{\small \it West University of Timi\c soara,}\\
{\small \it V. Parvan Ave. 4 RO-300223 Timi\c soara,  Romania}}

\begin{abstract}
In this letter we study the problem of scalar particle production in external electric field in de Sitter geometry. The total probability is calculated using the previously obtained result in Ref.\cite{6} for transition amplitude in external electric field on de Sitter space.  Then we make a graphical study of the total probability in terms of the ratio mass of the particle/expansion factor. Our results show that the probability depend of the direction in which the particles are emitted and that the probability becomes maxim when particles are emitted on the direction of the electric field. In the Minkowski limit we obtain that the probability is vanishing.
\end{abstract}


\maketitle

\section{Introduction}
The problem of particle production in electric field on de Sitter geometry received a lot of attention in the literature \cite{6,17,18,24}. This problem can be studied using nonperturbative methods \cite{1,7,8,9,12,13,17,18,21,22} and the main results are related to the number density of produced particles. More recent studies suggest that this effect could also appear as a perturbative phenomenon \cite{3,6,14,15,16,19,20,23,24}. This is because in de Sitter QED these processes are no longer forbidden by the simultaneous energy-momentum conservation as in flat space case. The cosmological mechanism of particle creation based on nonperturbative methods use the approximation of the out field at late times to obtain the density number of produced particles. In the case of scalar pair production it was shown that the results are valid only when the mass of the particle is larger than the expansion parameter $m>>\omega$ \cite{18,22}, and this corresponds to weak gravitational fields. The perturbative approach to the problem of particle production in de Sitter geometry, use the exact solutions of the free field equations for computing the transition probabilities. The results obtained so far using this approach show that these probabilities are nonvanishing only when the expansion factor is comparable (or larger) with the mass of the scalar particle, that corresponds to strong gravitational fields. In the limit $m>>\omega$, these probabilities are very small or even vanish since this limit is close to the flat space case. So it is clear from the start that a direct comparation between this two different mechanisms of particle production needs to be considered with care.

In this paper we use the result obtained in \cite{6}, for computing the total probability for the process of scalar pair production in external field. Our results prove that the total probability is nonvanishing only when $m\sim\omega$ and vanish in the Minkowski limit. Further we will make a series of observations related to a recent work \cite{13}, that compare the perturbative and nonperturbative methods for obtaining particle production in external field. Our analysis show that the results obtained in \cite{13}, cannot be used for comparing the density number obtained in \cite{6} with the existing nonperturbative results.

The letter is organized as follows: in the second section we present the results obtained for pair production in electric field using the perturbative approach. In the third section the total probability is computed and in section four we make a graphical analysis of the total probability. Section five is dedicated to the comparation between perturbative and nonperturbative approaches to the problem of particle production on de Sitter geometry. In section six we present our conclusions and in Appendix we give the mathematical formulas that help us to establish the final results for total probability.

\section{The transition amplitude}
In this section we want to summarize the results obtained in Ref.\cite{6}, where the production of the scalar particles in electric field on the de Sitter expanding universe was studied. In Ref.\cite{6} the calculations were performed using a perturbative approach.\\
In order to obtain the expression of the transition amplitude we define the $in$ and $out$ fields as being the exact solutions of the Klein-Gordon equation in momentum basis \cite{1,25}:
\begin{equation}\label{eq:sol}
f_{\vec{
p}}\,(x)=\frac{1}{2}\sqrt{\frac{\pi}{\omega}}\,\frac{e^{-3\omega
t/2}}{(2\pi)^{3/2}}\,e^{-\pi k/2}H^{(1)}_{ik}\left(\frac{p}{\omega}\,e^{-\omega t}\right) e^{i \vec{p}\cdot\vec{
x}}\,,
\end{equation}
where $H^{1}_{\mu}(z)$ is the Hankel function of first kind, $p=|\vec{p}|$. In addition we note:
\begin{equation}\label{eq:k}
\mu=\sqrt{k^2-\textstyle{\frac{9}{4}}}\,, \quad k=\frac{m}{\omega}\,,
\end{equation}
with $m>3\omega/2$.

We recall that in \cite{6} the external electric field $\vec{E}=\frac{Q}{|\vec{x}|^{2}}e^{-2\omega t}\vec{n}$ was produced by the vector potential $\vec{A}$:
\begin{eqnarray} \label{c1}
A^i=\frac{Q}{2\omega|\vec{x}|^{2}}e^{-2\omega t}n^i\,,
\end{eqnarray}
where the contravariant component of the four vector potential was used, since the scalar QED is constructed with vectors.

The final result for the transition amplitude obtained in Ref.\cite{6}, is expressed in terms of unit step function $\theta$ and gamma Euler function $\Gamma$ :
\begin{eqnarray}\label{re1}
S_{\varphi\,\varphi^{+}}=\frac{ie^2}{16\pi^{2}m}\frac{(\vec{p}-\vec{p}\,')\cdot\vec{n}}{|\vec{p}+\vec{p}\,'|}\left[\frac{1}{p^{2}}\theta(p-p\,')f_{\mu}\left(
\frac{p\,'}{p}\right)+\frac{1}{p\,'^{2}}\theta(p\,'-p)f_{\mu}\left(\frac{p}{p\,'}\right)\right],
\end{eqnarray}
where the function $f_{\mu}\left(\frac{p\,'}{p}\right)$ was defined as \cite{6}:
\begin{eqnarray}
f_{\mu}\left(\frac{p\,'}{p}\right)= \frac{k}{\left(1-\left(\frac{p\,'}{p}\right)^2\right)}\left[\left(\frac{p\,'}{p}\right)^{i\mu}\Gamma(1+i\mu)\Gamma(-i\mu)+
\left(\frac{p\,'}{p}\right)^{-i\mu}\Gamma(1-i\mu)\Gamma(i\mu)\right].
\end{eqnarray}
We must mention that $f_{\mu}\left(\frac{p}{p\,'}\right)$ is obtained when $p\rightleftarrows p\,'$ .
The result obtained in \cite {6} show that the momentum conservation law is broken in this process. This is  a direct consequence of the fact that the vector potential $\vec{A}=\vec{A}(t,x)$ is a function of the spatial coordinate $x$, resulting that the spatial integral gives a result which is not dependent of the delta Dirac function. The presence of the external field (\ref{c1}) in de Sitter geometry leads to the breaking of the momentum conservation law as was shown in \cite{24}.

\section{The total probability}

In this section we will present the main steps for calculating the total probability of scalar pair production in electric field on de Sitter space. For presenting our arguments we restrict only to the case when $p>p\,'$ with the observation that we consider the ratio of the momenta close but not equal to unity such that $p/p\,'\in(0.2,0.9)$. Using equations (\ref{p}) and (\ref{ln}) from Appendix, the $f_{\mu}$ functions that define the probability can be brought in this case to the form:
\begin{equation}\label{fa}
f_{\mu}\left(\frac{p\,'}{p}\right)=\frac{2\pi k\mu }{\sinh(\pi \mu)}\frac{p}{(p+p\,')}.
\end{equation}
The total probability is obtained by integrating after the final momenta $p\,,p\,'$ the probability density . Since the particles are emitted in pairs we will study the situation when the momenta $\vec{p}\,,\vec{p}\,'$ are emitted on the same direction with the direction of the electric field, which is given by the unit vector $\vec{n}$. The total probability will also be computed in the case when the particles will be emitted in the directions which do not coincide with the direction of the electric field. The nominator can be expressed in terms of the angle between the momenta vectors $\theta_{pp\,'}$:
\begin{equation}
|\vec{p}+\vec{p}\,'|^{2}=p^2+p\,'^2+2p\,p\,'\cos\theta_{pp\,'}
\end{equation}

Using the equation (\ref{re1}), the expression for probability density becomes ($p>p\,'$):
\begin{equation}\label{po}
P_{\varphi\varphi^+}=\frac{e^4\mu^2k^2}{64\pi^2m^2\sinh^2(\pi k)}
\,\frac{(p\, \cos(\theta_{pn})-p\,'\cos(\theta_{p\,'n}))^2}{p^2\,(p+p\,')^2\,(p^2+p\,'^2+2pp\,'\cos(\theta_{pp\,'}))},
\end{equation}
where $\theta_{p\,'n},\,\theta_{pn}$ are the angles between momenta vectors and the vector of electric field.
Then the total probability can be obtained by integrating after the final momenta:
\begin{equation}\label{tot}
P_{tot}=\int\int P_{\varphi\varphi^+}\,\,d^3p\,\,d^3p\,'
\end{equation}
We analyse only the case $p>p\,'$ since for $p\,'>p$ the calculations are similar. The integration after momenta modulus is taken such that $p\in(0,\infty)$ while $p\,'\in(0,p\,'_{E})$, where $p\,'_{E}=\sqrt{m^2+\frac{e^2E^2}{\omega^2}}$ is the particle momentum in the presence of the electric field $\vec{E}$. This cutoff of the momentum is also used in \cite{17,18} for avoiding divergent results for the number of particles. Here we adopt the same cutoff since we study the case when $p>p\,'$. A few important observations need to be made now. In the case of a perturbative calculation the probability is a function dependent of the particle momenta (see Eq.(\ref{po})) and in this case one may also apply other regularization methods for obtaining the total probability. In the existing nonperturbative results this is no longer valid since the approximations made for the out state give number densities which are not dependent on the particle momenta and the only remaining choice is to make a cutoff for the upper limits of the momenta \cite{17,18}.

First we analyse the case when the momenta of the pair have the same orientation $\theta_{pp\,'}=0$ and are emitted on the direction of the electric field $\theta_{pn}=\theta_{p\,'n}=0$. In other words we will study the total probability when the particles are emitted on given directions and for that reason it will be sufficiently to solve only the momenta modulus integrals.
Using now equations (\ref{fa}) and integrating after the final momenta $p\,,p\,'$ the total probability in the case $\theta_{pn}=\theta_{p\,'n}=\theta_{pp\,'}=0$ gives:
\begin{equation}\label{pt1}
P_{tot}=\frac{e^4\mu^2}{64\sinh^2(\pi\mu)}\,\frac{8}{3}\left(\left(\frac{m}{\omega}\right)^2+\frac{e^2E^2}{\omega^4}\right).
\end{equation}

Further we will compute the total probability in two distinct cases when the particles are emitted in other directions which make various angles with the direction of the electric field given by $\vec{n}$.
The final results in these cases turns to be:
\begin{equation}\label{pt2}
P_{tot}=\frac{e^4\mu^2}{64\sinh^2(\pi\mu)}\left\{
\begin{array}{cll}
\frac{5}{2}\left(\left(\frac{m}{\omega}\right)^2+\frac{e^2E^2}{\omega^4}\right)&{\rm for}&\theta_{pn}=0\,;\,\theta_{p\,'n}=\theta_{pp\,'}=\pi/6\\
\frac{7}{4}\left(\left(\frac{m}{\omega}\right)^2+\frac{e^2E^2}{\omega^4}\right)&{\rm for}&\theta_{pn}=\pi/3\,;\,\theta_{p\,'n}=\theta_{pp\,'}=\pi/6
\end{array}\right.
\end{equation}
The total probability depends on the parameter $k=m/\omega$, and from our analytical expression one can observe that this function is highly convergent for large $k$, since is proportional with a factor $\sinh^{-2}(\pi\mu)$.
From the above equation we observe that the processes in which the particles are emitted on the direction of the electric field are favoured since in this case the probability is larger (see Eqs.(\ref{pt1}),(\ref{pt2})). This result proves that there is a net distinction between the situations when the particles are emitted on the direction of electric field or when the particles are emitted in other directions. In \cite{13} the density number obtained using nonperturbative method depend on the orientation of the particle momenta relatively to the field directions. The same conclusion can be reached if we use perturbative methods as can be seen from our exact perturbative results obtained in Eqs.(\ref{pt1})-(\ref{pt2}). Our results do not sustain the conclusion reached in \cite{13} about the perturbative result and prove that the probability depend on the direction and momenta orientation of the particles, relatively to the electric field direction. Moreover in the perturbative case one can study the interesting situation when the particles are emitted on the field direction but their momenta have opposite orientation. This situation is not analysed in \cite{13} because there is no analytical expression for transition probability since the calculations were done numerically. A closer look at the perturbative result obtained in \cite{13} leave some questions. Let us recall the amplitude result, using the external field given in \cite{18}, $A_{x}=\frac{E_{o}}{\omega^2 t_{c}}$, where $t_{c}=-\frac{e^{-\omega t}}{\omega}$ is the conformal time and $E_{o}$ is the intensity of the electric field. Then the amplitude obtained in \cite{13} without solving the temporal integrals is proportional with:
\begin{eqnarray}
&&A_{i\rightarrow f}\sim \,\delta^{3}(\vec{p}+\vec{p}\,')\int_{0}^{\infty}dz
H_{ik}^{(2)}(pz)H_{ik}^{(2)}(p\,'z),
\end{eqnarray}
where the variable of integration is $z=-t_{c}$.
From the delta Dirac term $\delta^{3}(\vec{p}+\vec{p}\,')$ it is immediate that $\vec{p}=-\vec{p}\,'$ that means that the momenta are equal in modulus $p=p\,'$. The problem now is related to the temporal integral which contain in general the result when $p\neq p\,'$ and a direct computation of this integral prove that the result can be expressed in terms of Gauss hypergeometric functions that have the algebraic argument $z=\frac{p}{p\,'}$, which can not be equal with $1$, since in these conditions these functions become divergent. By solving numerically the integral in the case $p=p\,'$ one needs to cut the upper limit of the temporal integral. This means that the upper limit of the variable $z$ is finite, which also lead to questions regarding the justification for this cutoff. So an analytical study of these integrals is required for a correct and well justified result.

Another observation about to the result obtained in \cite{13} is related to the justification of the result in two dimensions. The interaction between photons and charged scalar particles it is known in physics as the scalar quantum electrodynamics. This theory is constructed in four dimensions and for that we have a good reason: there is no electromagnetic field in one spatial dimension since the photons have polarization $\vec{\varepsilon}_{\lambda} (\vec{k\,})$, which is perpendicular on the photon momenta $\vec{k}\cdot\vec{\varepsilon}_{\lambda} (\vec{k\,})=0$ \cite{4,5,23}(the field is transversal). This is why it is impossible to calculate perturbative amplitudes in two dimensions (one spatial dimension and one temporal dimension) when one study interactions intermediate by photons, or to reduce the result obtained in four dimensions to two dimensions. In one spatial dimension one could not construct the theory of electromagnetic field and further the reduction formalism and interaction theory. All the QED models in two dimensions have to be considered with care since the they work with massless fields and the interaction lagrangean will contain additional terms, which will change the definition of the transition amplitudes. All these aspects should be clear from the Minkowski QED \cite{4,5} and the same observations apply for de Sitter QED \cite{3,23,24}. So the amplitude from \cite{13} computed in two dimensions should be considered with care since as we explained above there is no perturbative QED in two dimensions. In fact the perturbative methods \cite{3,6,23,24} seems to be very good in describing what happens at large expansion $\omega>> m$, with the observation that in this case the nonperturbative approach becomes problematic in the case of a scalar field as was shown in \cite{18,22}.

\section{Graphical results}

Further we plot the total probability as function of parameter $k=m/\omega$, giving different numerical values to the parameter $e^2E^2/\omega^4$. We present in our plots the situation when the particles are emitted on the direction of electric field and the situation when the pair is emitted in other directions which do not coincide with the direction of the electric field.

\begin{figure}[h!t]
\includegraphics[scale=0.42]{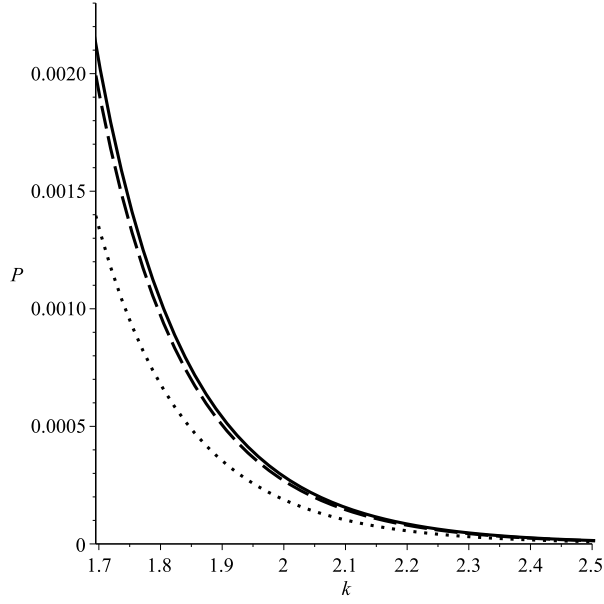}
\caption{The total probability dependence of $k=m/\omega$,$\frac{e^2E^2}{\omega^4}=0.001$. The solid line is for $\theta_{pn}=\theta_{p\,'n}=\theta_{pp\,'}=0$ , the dashed line for $\theta_{pn}=0\,;\,\theta_{p\,'n}=\theta_{pp\,'}=\pi/6$ and the point line for
$\theta_{pn}=\pi/3\,;\,\theta_{p\,'n}=\theta_{pp\,'}=\pi/6$}
\label{fg1}
\end{figure}

\begin{figure}[h!t]
\includegraphics[scale=0.42]{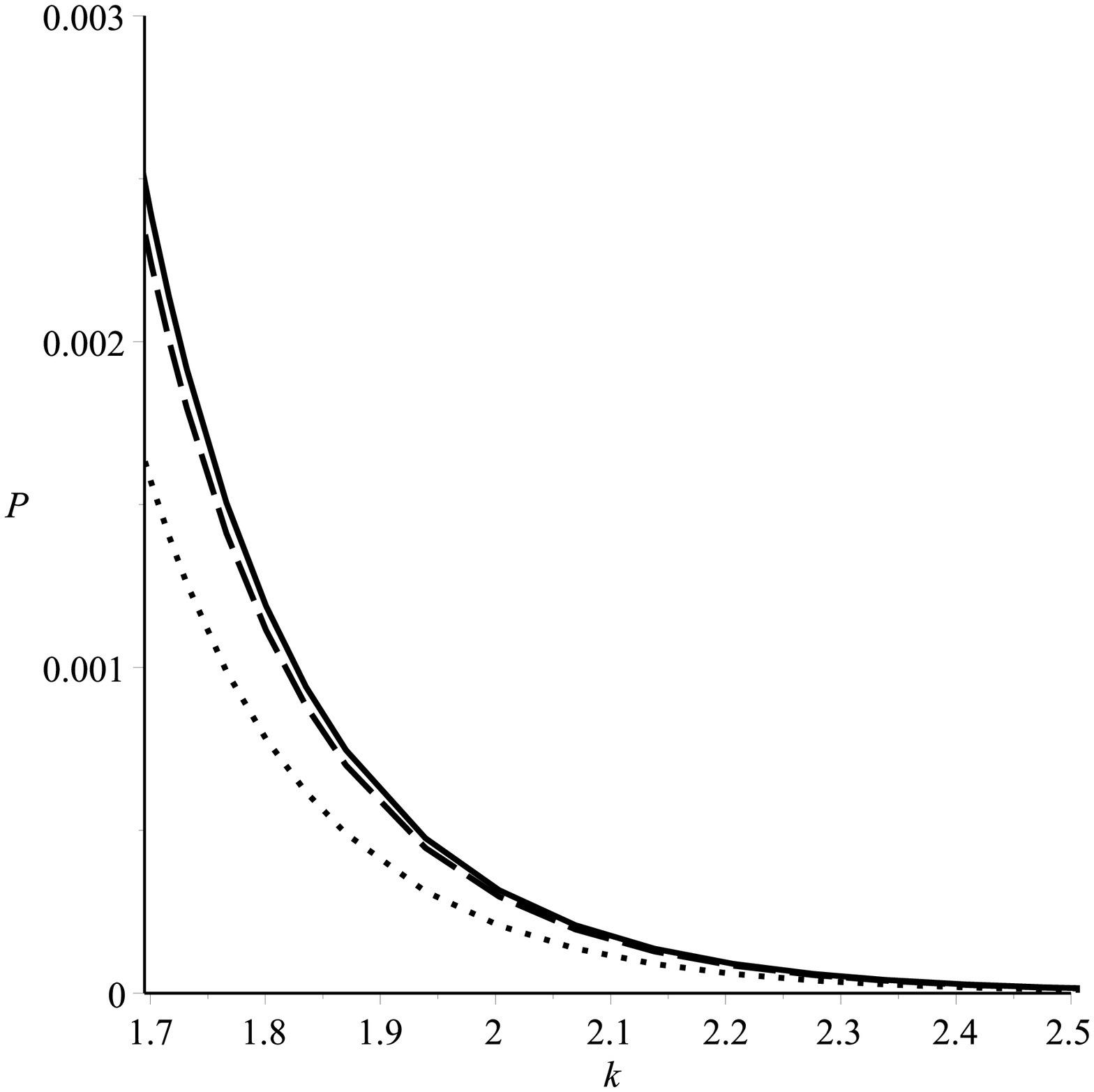}
\caption{The total probability dependence of $k=m/\omega$,$\frac{e^2E^2}{\omega^4}=0.5$. The solid line is for $\theta_{pn}=\theta_{p\,'n}=\theta_{pp\,'}=0$ , the dashed line for $\theta_{pn}=0\,;\,\theta_{p\,'n}=\theta_{pp\,'}=\pi/6$ and the point line for
$\theta_{pn}=\pi/3\,;\,\theta_{p\,'n}=\theta_{pp\,'}=\pi/6$}
\label{fg2}
\end{figure}

\begin{figure}[h!t]
\includegraphics[scale=0.42]{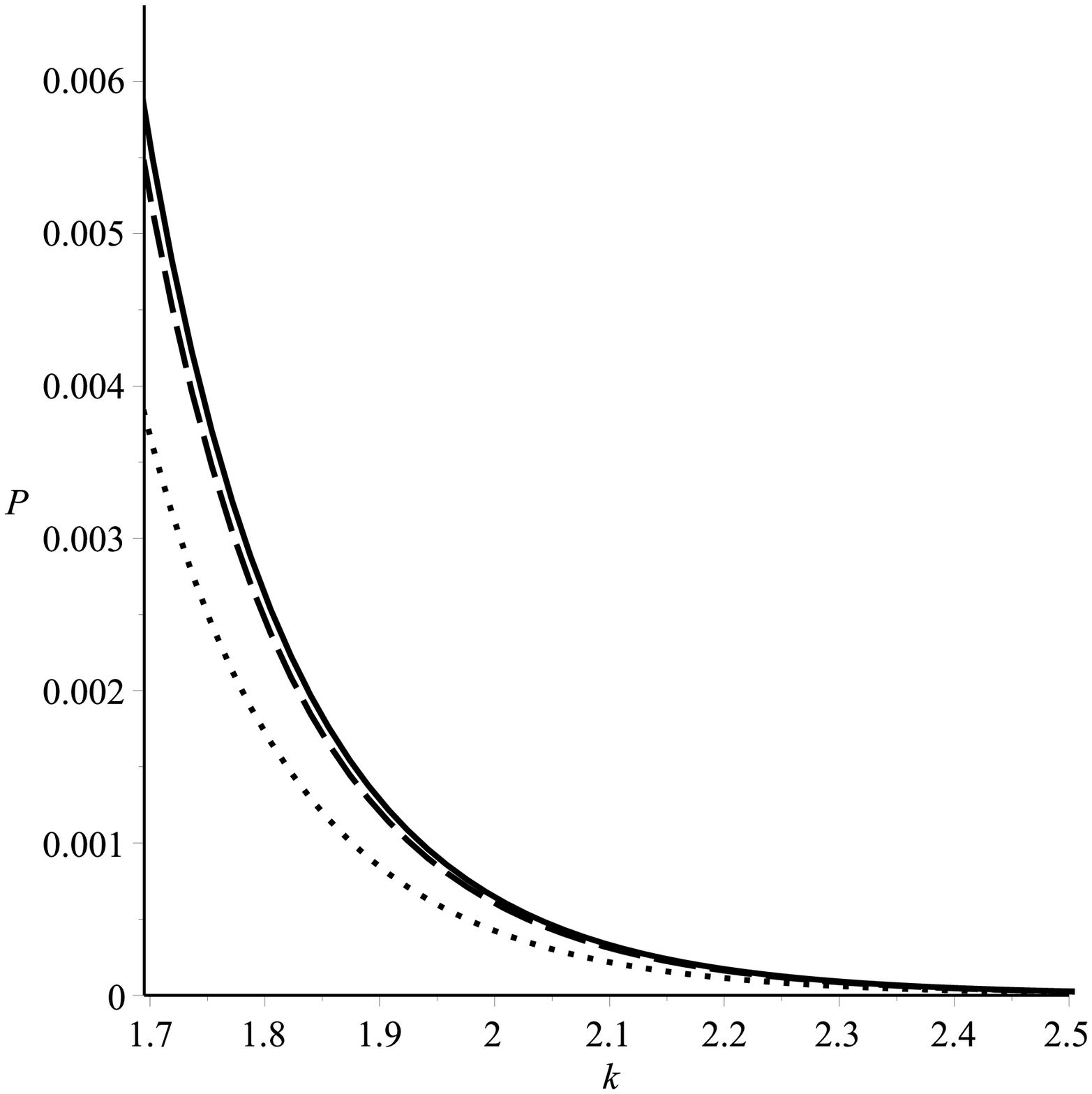}
\caption{The total probability dependence of $k=m/\omega$,$\frac{e^2E^2}{\omega^4}=5$. The solid line is for $\theta_{pn}=\theta_{p\,'n}=\theta_{pp\,'}=0$ , the dashed line for $\theta_{pn}=0\,;\,\theta_{p\,'n}=\theta_{pp\,'}=\pi/6$ and the point line for
$\theta_{pn}=\pi/3\,;\,\theta_{p\,'n}=\theta_{pp\,'}=\pi/6$}
\label{fg3}
\end{figure}

\begin{figure}[h!t]
\includegraphics[scale=0.42]{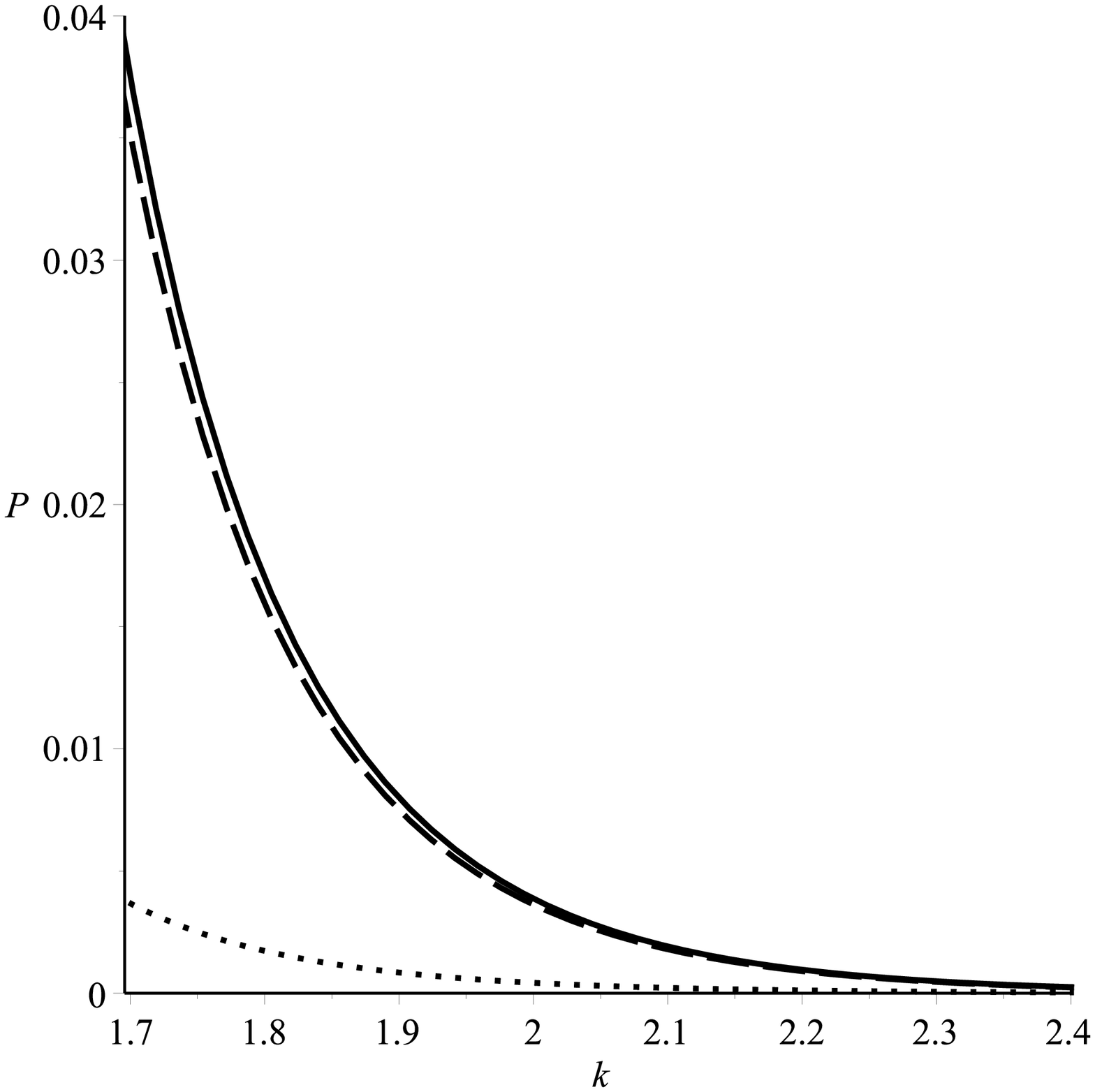}
\caption{The total probability dependence of $k=m/\omega$,$\frac{e^2E^2}{\omega^4}=50$. The solid line is for $\theta_{pn}=\theta_{p\,'n}=\theta_{pp\,'}=0$ , the dashed line for $\theta_{pn}=0\,;\,\theta_{p\,'n}=\theta_{pp\,'}=\pi/6$ and the point line for
$\theta_{pn}=\pi/3\,;\,\theta_{p\,'n}=\theta_{pp\,'}=\pi/6$}
\label{fg4}
\end{figure}

From Figs.(\ref{fg1})-(\ref{fg4}), we observe that the total probability drops rapidly to zero as the parameter $m/\omega\sim2$ and this behaviour is also preserved when we increase the intensity of the electric field (see Figs.\ref{fg3}-\ref{fg4}). Indeed the probability becomes larger as we increase the parameter $e^2E^2/\omega^4$, but have nonvanishing values only when $m\sim \omega$. The conclusion is that the perturbative result is nonvanishing only when the electric field is coupled with a strong gravitational field and vanish for $m>>\omega$. This observation is important since the nonperturbative results for computing the beta Bogoliubov coefficient are obtained approximating the Klein-Gordon equation for $m>>\omega $. It is well known that the Schwinger effect appear in the Minkowski theory as a nonperturbative phenomenon and in de Sitter space applying the same nonperturbative methods this effect could also be obtained \cite{17,18}. Returning now to the de Sitter QED it was proved that any external field coupled with the strong gravity of the early Universe will give nonvanishing probabilities for the QED processes in the first order of perturbation theory \cite{3,6,24}, which clearly must reduce to zero in the Minkowski limit as the laws of energy-momentum conservation ask. So as long as one use the perturbative methods to investigate pair production in electric fields on de Sitter geometry, we do not expect to recover the Schwinger effect in the Minkowski limit. In \cite{13}, the graphycal results show (using a numerically evaluation of the integral) the number of particles in terms of $eE/\omega^2$ for given values of $m/\omega$, without taking into account the  above observations and compare the results of the perturbative method and nonperturbative method, for the same values of the parameter $m/\omega$. In our opinion this is is not the right think to do since the two methods give good results for completely different rates of space expansion. This is because the Bogoliubov coefficients are computed approximating the solutions of the free field equation in the limit $m>>\omega$ \cite{17,18}. Another problem  worth to be mentioned is that the nonperturbative calculations are performed in two dimensions, since in four dimensions there are serious technical problems for completing the calculations. For finding some similarities between the results from \cite{17,18} and the result obtained in \cite{6} one needs to take the potential given in Eq.(\ref{c1}) and solve the Klein-Gordon equation in four dimensions in the case $m>>\omega$, which is not at all an easy task. Then approximate the out state such that the Bogoliubov coefficients could be determined. The last step requires to obtain an analytical formula for the perturbative amplitude at $m>>\omega$, which is very small but not zero. In the perturbative case this result can be obtained and the probability is proportional in this case with a factor $e^{-2\pi\sqrt{(m/\omega)^2-9/4}}$. Even so the results will be different since the methods are of completely different nature.

To be more clear let us consider the pair production amplitudes calculated with the field produced by a point charge (or a distribution of point charges) in Minkowski QED, which are zero. In contrast if we study the same problem in de Sitter QED \cite{6}, the result is nonvanishing only at large expansion. So we study the problem of pair production in weak electric fields coupled with the strong gravity (large expansion $\omega\sim m$) of the early Universe. This problem have nothing in common with the Schwinger effect which is obtained using a nonperturbative approach in Minkowski space \cite{0}. This result can be obtained using the same nonperturbative approach in de Sitter space \cite{17,18}, when $m>>\omega$. Our results refers to the case when the electric fields are weak and gravity becomes strong and in the end we recover the Minkowski limit which is zero as we expect. In the nonperturbative approach the situation is inverse since the electric fields are strong and the equations are studied at $m>>\omega$, which clearly means weak gravity. Finally we conclude that the result obtained in \cite{13} do not take into consideration the above observations and in addition there are problems in considering QED calculations in two dimensions as we mentioned in the previous section.

\section{Perturbative/nonperturbative approach}

In this section we will make a series of observations related to the perturbative and nonperturbative approach to the problem of pair production in external field on de Sitter geometry. First we want to comment some of the affirmations made in \cite{13}, about the result obtained in \cite{6}. In the introduction of \cite{13} it was suggested that in Ref.\cite{6}, was obtained the number of particles using a perturbative approach. In \cite{6} only the probability density was computed and further integration after the final momenta needs to be done for obtaining the particle number. In fact in the perturbative case the problem of defining the particle number is not well studied. Another observation is that in \cite{13} only the nonperturbative results about the problem of pair production are cited giving the impression that only Ref.\cite {6} use perturbative methods. In fact there is an extended literature that use perturbative methods \cite{3,14,15,16,19,20,24,fo}, and it is only fair to mention these papers. The problem of particle production due to the fields interactions was also studied in \cite{12,fo}. In \cite{12,fo} it was proven that this kind of particle production could become important and must be taken into account along side with the cosmological particle production. In the case when one use external fields as those given in \cite{6} the spatial integrals and temporal integrals give completely different results. These results show that the momentum is no longer conserved for pair production in external fields, which have a dependence of spatial coordinates. Moreover the perturbative probability depends on the ratio $m/\omega$ and on the particle momenta $p,\,p\,'$ as one can see from equation (\ref{po}). When the total probability is computed the integrand will depend on the momenta and in the case of divergent integrals one can apply regularization methods. In the nonperturbative case the approximations done for the out modes have as consequence the lose of the momenta dependence in the beta $\beta$ Bogolibov coefficient. Then an integration over the final momenta will give a quantity which is at least linearly divergent depending on the number of dimensions used in the problem. For that reason a cutoff of the upper limits of the momenta integration need to be done as we mentioned in the previous section. Even so one can justify these cutoffs and important results were obtained in \cite{7,8,9,17,18,22}. But in the present paper we do not want to criticise the nonperturbative method which gives important results as we mentioned above. Instead we want to point out that both approaches to the problem of particle production must be used in order to obtain a clear picture about the phenomenon of particle production in gravitational fields. One cannot neglect the pair production processes that arise when the perturbative methods are used, which do not contradict the existing nonperturbative results \cite{7,8,9,17,18,21,22}. The results obtained so far using perturbative methods prove that some physical consequences related to the conservation laws and the mechanism of separation between matter and antimatter could be studied and this method is more suitable for analysing the phenomenon of pair production in strong gravitational fields \cite{3,6,7,8,9,14,15,16,24}.

An interesting observation can be made related to the results obtained by this two methods. In the limit $m>>\omega$ the probabilities computed with perturbations is proportional with a term of the form $e^{-2\pi\sqrt{(m/\omega)^2-9/4}}$. The beta Bogolibov coefficient is also proportional with a term of this type. So for $m>>\omega$ the perturbative results reproduce up to some factors (which depend on the particle momenta), the nonperturbative results but this is not an argument for comparing two different mechanisms of particle production.

The mechanism of cosmological particle production use the fact that the vacuum could become unstable and for that reason the solutions of the free field equations of two different local charts could be related using Bogoliubov transformations. These transformations mix the solutions which describe the particle with the solutions which describe the antiparticle. In this case the solutions of the field equations are defined locally and depend on the local chart, which means that the "in" and "out" modes are not the same. These effects could be observed using local particle detectors. This is the nonperturbative approach to the problem of particle production \cite{21}. If one use the perturbative method then the modes are globally defined on entirely manifold and do not depend on local coordinates providing that the vacuum state is stable and unique \cite{3}. In this case the quantum states are measured using a global apparatus which consist of conserved operators. Then the transition probabilities can be calculated using the exact solutions of the free field equations. In the scalar QED on de Sitter geometry we can study the effect of electromagnetic interaction on the particle production \cite{6}. These effects are present only when the gravitational field is still strong ($m\sim\omega$) and vanish in the limit $m>>\omega$. So the two mechanisms that study the problem of particle production are of completely different nature and it is premature to draw definitive conclusions in the absence of experimental evidence or to criticize the existing perturbative results without having strong mathematical and physical arguments \cite{13}.

\section{Conclusions}
In this paper we study the total probability for massive scalar particle production in electric field on de Sitter geometry. Our results prove that the probability depend on the parameter $m/\omega$ and vanish in the Minkowski limit. This result is expected since we consider here the field produced by a distribution of point charges in de Sitter geometry. Our study proves that the probability is nonvanishing only for $m\sim\omega$ regardless of the intensity of the electric field and the probability varies with the directions in which particles are emitted. These conclusions are confirmed also by our graphical analysis. When the electric field is strong we obtain that the probability of particle emission is larger when the particles are emitted in the direction of the electric field.

Our results also prove that a direct comparison between nonperturbative methods and perturbative methods in what concerns the phenomenon of particle production, must be considered with care since the two mechanisms are completely different.

\section{Appendix}
For obtaining function $f_{\mu}\left(\frac{p\,'}{p}\right)$, we use:
\begin{equation}\label{p}
a^{ix}=e^{ix\ln a},\,\,a\in\emph{R}
\end{equation}
and
\begin{equation}\label{ln}
\ln y=y-1-\frac{(y-1)^2}{2}+\frac{(y-1)^3}{3}+...\,\,,\,\,|y-1|\leq1.
\end{equation}

The momenta integrals that help us to establish the final formulas for the total probability are:
\begin{eqnarray}\label{pn}
\int_{0}^{\infty}\frac{dp\,(p-\frac{\sqrt{3}}{2}p\,')^2}{(p+p\,')^2(p+p\,'+\sqrt{3}pp\,')}=\frac{1}{8p\,'(2-\sqrt{3})}(14+8\sqrt{3}-5\pi\sqrt{3}),\nonumber\\
\int_{0}^{\infty}\frac{dp\,(p-\sqrt{3}p\,')^2}{(p+p\,')^2(p+p\,'+\sqrt{3}pp\,')}=\frac{1}{3p\,'(2-\sqrt{3})}(12+6\sqrt{3}-4\pi\sqrt{3}),
\end{eqnarray}
\begin{equation}\label{ps}
\int_{0}^{\infty}dp\,\frac{(p-p\,')^2}{(p+p\,')^4}=\frac{1}{3p\,'}.
\end{equation}


\begin{thebibliography}{99}
\bibitem{0}
J. Schwinger , Phys. Rev. \textbf{82}, 664 (1951).
\bibitem{1}
E. Schr\" odinger,  Physica \textbf{6}, 899 (1939).
\bibitem{2}
 C.W.Misner, K.S.Thorne and J.A.Wheleer ,{\em Gravitation}
(W.H.Freeman and Company New York, 1973).
\bibitem{3}
I. I. Cot\u aescu and C. Crucean, {\em Phys. Rev. D} \textbf{87}, 044016 (2013).
\bibitem{4}
S. Weinberg, {\em The Quantum Theory of Fields}  (Cambridge University Press, Cambridge, 1995).
\bibitem{5}
 S.Drell and J.D.Bjorken, {\em Relativistic Quantum Fields}
(Mc Graw-Hill Book Co., New York 1965).
\bibitem{6}
M. A. B\u {a}loi, Mod. Phys. Lett. A \textbf{29}, 1450138, (2014).
\bibitem{7}
L.Parker, Phys.Rev.Lett. \textbf{21}, 562 (1968).
\bibitem{8}
L.Parker, Phys.Rev. \textbf{183}, 1057 (1969).
\bibitem{9}
L.Parker, Phys.Rev.D \textbf{3}, 346 (1971).
\bibitem{fo}
N. D. Birrel, P. C. W. Davies and L. H. Ford, {\em J. Phys. A} \textbf{13}, 961 (1980).
\bibitem{12}
 N. D. Birrel and P. C. W. Davies,  {\em Quantum Fields in Curved Space}
(Cambridge University Press, Cambridge 1982).
\bibitem{13}
N. Nicolaevici, Mod. Phys. Lett. A \textbf{30}, 1550046   (2015).
\bibitem{14}
K.H.Lotze, Nucl.Phys.B \textbf{312}, 673 (1989).
\bibitem{15}
K.H.Lotze, Class.Quant.Grav. \textbf{2}, 351 (1988).
\bibitem{16}
K.H.Lotze, Class.Quant.Grav. \textbf{5}, 595 (1985).
\bibitem{17}
V.M.Villalba, Phys.Rev.D \textbf{52}, 3742 (1995).
\bibitem{18}
J.Garriga, Phys.Rev.D \textbf{49}, 6343 (1994).
\bibitem{19}
I.L. Buchbinder, E.S. Fradkin and D.M. Gitman, Fortschr. Phys. \textbf{29}, 187 (1981).
\bibitem{20}
I. L. Buchbinder and L. I. Tsaregorodtsev, Int. J. Mod. Phys. A \textbf{7}, 2055 (1992).
\bibitem{21}
G. W. Gibbons and S. W. Hawking, Phys. Rev. D \textbf{15}, 2738 (1977).
\bibitem{22}
J.Haro and E.Elizalde, J.Phys. A \textbf{41}, 372003 (2008).
\bibitem{23}
I. I. Cot\u aescu and C. Crucean , Prog.T.Phys \textbf{124}, 1051 (2010).
\bibitem{24}
C. Crucean, Phys. Rev. D \textbf{85}, 084036 (2012).
\bibitem{25}
 I. I. Cot\u aescu, C. Crucean, A. Pop, Int. J. Mod. Phys.A {\bf 23} (2008) 2563.
\bibitem{26}
C. Crucean and M. A. B\u{a}loi, Int. J. Mod. Phys. A \textbf{30}, 1550088 (2015).
\end{thebibliography}
\end{document}